\documentclass[11pt,letterpaper]{article}%
\usepackage[usenames,dvipsnames,svgnames,table]{xcolor}
\usepackage{color}
\usepackage{amsmath}
\usepackage{amsfonts}
\usepackage{verbatim}
\usepackage{amssymb}
\usepackage{graphicx}
\usepackage{epstopdf}
\usepackage{mathrsfs}%
\providecommand{\U}[1]{\protect\rule{.1in}{.1in}}
\begin{document}

\title{Mass of asymptotically anti-de Sitter hairy spacetimes}

\author{$^{(1)}$Andr\'{e}s Anabal\'{o}n, $^{(2,3)}$Dumitru Astefanesei and $^{(4,5)}$Cristi\'an Mart\'{\i}nez\\  [1mm]
\textit{ \small $^{(1)}$Departamento de Ciencias, Facultad de Artes Liberales and}\\\textit{\small Facultad de Ingenier\'{\i}a y Ciencias, Universidad Adolfo
Ib\'{a}\~{n}ez,}\\\textit{ \small Av. Padre Hurtado 750, Vi\~{n}a del Mar, Chile}\\
\textit{ \small $^{(2)}$Instituto de F\'\i sica, Pontificia Universidad Cat\'olica de
Valpara\'\i so, Casilla 4059, Valpara\'{\i}so, Chile}\\
\textit{ \small $^{(3)}$ Max-Planck-Institut f$\ddot{u}$r Gravitationsphysik, Albert-Einstein-Institut, 14476 Golm, Germany}\\
\textit{\small  $^{(4)}$Centro de Estudios Cient\'{\i}ficos (CECs), Casilla 1469, Valdivia, Chile}\\
\textit{ \small $^{(5)}$Universidad Andr\'es Bello, Av.\ Rep\'{u}blica 440, Santiago, Chile}\\
[1mm]
\tt{\small andres.anabalon@uai.cl, dumitru.astefanesei@ucv.cl, martinez@cecs.cl}}
\maketitle

\begin{abstract}
In the standard asymptotic expansion of four dimensional static
asymptotically flat spacetimes, the coefficient of the first subleading term
of the lapse function can be identified with the mass of the spacetime. Using 
the Hamiltonian formalism we show that, in asymptotically locally anti-de Sitter 
spacetimes endowed with a scalar field, the mass can read off in the same 
way only when the boundary conditions are compatible with the asymptotic 
realization of the anti-de Sitter symmetry. In particular, this implies that some 
prescriptions for computing the mass of a hairy spacetime are not suitable when 
the scalar field breaks the asymptotic anti-de Sitter invariance.

\end{abstract}

\section{Introduction}

Scalar fields play a significant role in physics. From a theoretical
point of view, they are expected to be amongst the basic
constituents of fundamental theories, e.g. string theory.
Cosmologically, they are at the basis of inflation and dark energy
models. More importantly, the first fundamental scalar particle was
experimentally discovered \cite{Aad:2012tfa, Chatrchyan:2012ufa}. In
the last years, phenomenological applications of the physics of
hairy black holes have been proposed in different contexts. For
example, some of these configurations have found interesting
applications in condensed matter by using gauge/gravity dualities
(for a review see for instance \cite{Horowitz:2010gk} ). 
Additionally, astrophysical black holes have received a growing attention 
following the advent of new observational facilities and, consequently, 
different measurements for testing the spacetime geometry around 
these objects have been proposed. In particular, a cornerstone is to 
test the no-hair theorem from observations, i.e.  whether or not the 
black hole at the center of our galaxy belongs to the Kerr 
class (see for instance \cite{Will:2007pp, Loeb:2013lfa}). Therefore, \textit{ exact} hairy
black hole solutions have an essential role in conjunction with adequate formalisms 
to determine their physical properties, such as the mass and 
angular momentum, even in the presence of matter fields.

Such hairy configurations were ruled out by no-hair theorems for
which an asymptotically flat behavior for the gravitational field
and positivity of the scalar field potential are assumed
\cite{Chase,Bekenstein:1971hc,Teitelboim:1972qx,Bekenstein:1995un,Sudarsky:1995zg,Heusler:1996ft}. However,
the presence of a negative cosmological constant allows to
circumvent these theorems in a physically sensible way. Indeed, a number of
exact asymptotically anti-de Sitter (AdS) scalar hairy black holes have been
obtained following the precursor ones in three \cite{Martinez:1996gn,Henneaux:2002wm} 
and four dimensions  \cite{Martinez:2004nb}.
Recently, general classes of exact static hairy black hole solutions
have been obtained \cite{Anabalon:2012ta, Acena:2012mr,
Anabalon:2013qua, Acena:2013jya, Feng:2013tza, Anabalon:2013sra}, as
well as time dependent hairy black holes \cite{Zhang:2014sta,
Lu:2014eta}, which in turn has opened the possibility of
investigating their generic properties. For special values of the
parameters in the moduli potential, some of these solutions can
explicitly be embedded in supergravity theories \cite{Feng:2013tza,
Anabalon:2013eaa, Lu:2014fpa}.

An interesting physical effect emerges from asymptotically AdS scalar hairy 
solutions. Depending on its mass, the scalar field could acquire a slow
fall-off at infinity. In this case, the scalar field induces a strong back reaction 
on the metric and, in this sense, it cannot be treated as a probe. In particular, it 
was shown by using the Hamiltonian formalism that the scalar field contributes 
to the mass of the hairy 
solution \cite{Henneaux:2002wm, Henneaux:2004zi,Henneaux:2006hk} (other approaches 
 \cite{Barnich:2002pi,Gegenberg:2003jr,Hertog:2004dr,Banados:2005hm,Amsel:2006uf} have confirmed this result).

A relevant interval for the mass of the scalar field  where the above effect appears is
$m_{BF}^{2}\leq m^{2}<m_{BF}^{2}+l^{-2}$, where $m_{BF}^{2}=-9 l^{-2}/4$ is the
Breitenlohner-Freedman (BF) bound \cite{BF} and $l$ is the AdS radius. In this range 
the evolution of scalar fields in AdS is well defined for any linear combination of Dirichlet 
and Neumann boundary conditions  \cite{Ishibashi:2004wx}.

In the Hamiltonian formalism, the generators of the asymptotic
symmetries --- the conserved charges --- contain a bulk term that is a linear combination of the
constraints supplemented with a boundary term. The boundary term is fixed by requiring 
that the canonical generators have well-defined functional derivatives with respect to
the canonical variables \cite{Regge:1974zd}.  By virtue of the constraint equations, only the
boundary term contributes to the charges and so, from this
point of view, the Hamiltonian method is indeed suitable for a
holographic interpretation. Since the charges can be computed from the boundary term only, they
require just the asymptotic behaviour of the canonical variables and symmetries. Thus, the
canonical generators provide the charges for all the solutions sharing the same asymptotic behaviour. 

In this letter, we re-examine the notion of mass for asymptotically AdS scalar hairy
configurations in the framework of General Relativity with a minimally coupled scalar 
field. The tool, we are going to use for computing the mass,
is the Hamiltonian method of Regge and Teitelboim \cite{Regge:1974zd},  
following the results of \cite{Henneaux:2006hk}. 

A remarkable feature of this class of solutions is found. Once the canonical 
generator associated to the time translation (that corresponds from 
first principles to the mass of the configuration\footnote{Hereafter, we 
name it \textit{Hamiltonian mass} just for remarking its origin.}) is evaluated using the
equations of motions,  its value coincides with the coefficient of the first subleading term
of the lapse function only for boundary conditions that are compatible 
with the canonical realization of the local AdS symmetry at the boundary.

We would like to keep the
discussion concrete and, therefore, we treat the case of a single
scalar field with the conformal mass, $m^{2}=-2l^{-2}$, which is in
the allowed interval. In this case, both modes of the scalar
are normalizable. This value of the mass is relevant for gauged
supergravities in four dimensions \cite{de Wit:1982ig} and we can 
explicitly apply our general results to analytic hairy
black hole solutions \cite{Martinez:2004nb,Anabalon:2012ta, Acena:2013jya,
Anabalon:2013eaa}. This mass is also 
interesting because it allows for
subleading logarithmic branches (depending on the form of the scalar
field potentials and boundary conditions \cite{Henneaux:2004zi, Henneaux:2006hk}), which
needs to be treated separately. We expect that similar results should hold for
scalar fields with arbitrary mass in the interval $m_{BF}^{2}\leq m^{2}%
<m_{BF}^{2}+l^{-2}$ and for any dimensions.

There are  different proposals in the literature, developed from other 
rationale, for computing the mass. It is interesting to study the conditions
that enable those prescriptions to provide the right mass for the solutions 
analyzed here. In particular, the formula of
Ashtekar-Magnon-Das (AMD) \cite{Ashtekar:1984zz, Ashtekar:1999jx}
have been extensively used to obtain the mass of different hairy
configurations  \cite{Acena:2012mr, Acena:2013jya, Lu:2013ura,
Chow:2013gba, Liu:2013gja, Chen:2005zj}. We explicitly show that the
AMD mass matches the Hamiltonian mass of hairy configurations only
for boundary conditions that are compatible with 
the local AdS symmetry at the boundary.

\section{Hamiltonian mass}

Let us consider the action for a real scalar field minimally coupled to
four-dimensional Einstein gravity in the presence of a cosmological constant
$\Lambda=3 l^{-2}$ and a self-interaction potential $U(\phi)$
\begin{equation}
I[g_{\mu\nu},\phi]=\int d^{4}x\sqrt{-g}\left(  \frac{R-2\Lambda}{2\kappa
}-\frac{1}{2}g^{\mu\nu}\partial_{\mu}\phi\partial_{\nu}\phi-U(\phi)\right),
\label{eq:action}%
\end{equation}
where $\kappa=8\pi G$ is the Einstein constant. The corresponding field equations are
\begin{equation}
E_{~~\nu}^{\mu} \equiv G_{~~\nu}^{\mu}+\Lambda\delta_{~~\nu}^{\mu}-\kappa\left[  \partial^{\mu}%
\phi\partial_{\nu}\phi-\left(  \frac{1}{2}\partial^{\mu}\phi\partial_{\mu}%
\phi+U(\phi)\right)  \delta_{~~\nu}^{\mu}\right]=0,
\end{equation}
and
\begin{equation}
\Box\phi -\frac{dU(\phi)}{d\phi}=0.
\label{KG}%
\end{equation}

As we have mentioned before, we consider the Regge-Teitelboim approach to compute the mass of static
scalar hairy  asymptotically locally AdS  spacetimes. A summary of this method is provided below.
  
The canonical generator of an asymptotic symmetry defined by the vector $\xi=(\xi^{\perp},\xi^{i})$
is a linear combination of the constraints $\mathcal{H}_{\perp}, \mathcal{H}_{i}$ plus a surface term $Q[\xi]$ 
\begin{equation}
H[\xi]=\int d^{3} x \left(  \xi^{\perp} \mathcal{H}_{\perp}+\xi^{i}%
\mathcal{H}_{i}\right)  +Q[\xi].
\end{equation}
The surface term is chosen so that the generator has well-defined functional derivatives \cite{Regge:1974zd}.

In the static case the relevant asymptotic symmetry corresponds to the Killing
vector $\partial_{t}$, and the mass is the conserved charge associated with this
asymptotic symmetry. In the presence of the scalar field, apart from the usual 
gravitational term,  an extra term coming from it appears in the mass. We then
write the variation of the mass  \cite{Henneaux:2006hk} as
\begin{equation} \label{qt}
\delta M=\delta Q(\partial_{t})=\delta M_{G}+\delta M_{\phi},%
\end{equation}
where
\begin{equation}
\delta M_{G}(\xi)=\frac{1}{2\kappa}\int dS_{l}G^{ijkl}(\xi^{\bot}\delta
g_{ij;k}-\xi_{~,k}^{\bot}\delta g_{ij}),\label{eq:Q_G}%
\end{equation}
and
\begin{equation}
\delta M_{\phi}(\xi)=-\int dS_{l}\xi^{\bot}g^{1/2}g^{lj}\partial_{j}\phi
\delta\phi. \label{eq:Q_phi}%
\end{equation}
Here, $g_{ij}$ denotes the components of the $3$-spatial metric, $g=\det
g_{ij}$, and, as usual, we define
\begin{equation}
G^{ijkl}\equiv\frac{1}{2}g^{1/2}(g^{ik}g^{jl}+g^{il}g^{jk}-2g^{ij}g^{kl}).
\end{equation}

The variation of the mass given in (\ref{qt}) requires asymptotic boundary conditions
to be integrated. As is expected from physical grounds, the mass is well defined after imposing 
suitable boundary conditions.

\subsection{Non-logarithmic branch}

For a self-interaction potential, whose power series expansion around $\phi=0$
has the mass term $m^{2}=-2l^{-2}$ and a vanishing cubic term, the
asymptotically AdS behavior for the metric and scalar field does not contain
logarithmic branches \cite{Henneaux:2006hk}. Following this reference, we
consider a set of asymptotic conditions, that will be described below, and for
which there exist analytic scalar black hole solutions \cite{Martinez:2004nb,Anabalon:2012ta,
Acena:2013jya, Anabalon:2013eaa} whose  asymptotic behavior belong 
the chosen one. The fall-off of the scalar field is
\begin{equation}
\phi=\frac{\alpha}{r}+\frac{\beta}{r^{2}}+O(r^{-3}), \label{phi1}%
\end{equation}
where $\alpha$ and $\beta$ denote two real constants. For static metrics that
match (locally) AdS at infinity, the relevant fall-off is
\begin{align}
-g_{tt}  &  =\frac{r^{2}}{l^{2}}+k-\frac{\mu}{r}+O(r^{-2}),\label{BCA}\\
g_{mn}  &  =r^{2}h_{mn}+O(r^{-1}),\label{BCB}\\
g_{rr}  &  =\frac{l^{2}}{r^{2}}+\frac{al^{4}}{r^{4}}+\frac{l^{5}b}{r^{5}%
}+O(r^{-6}), \label{BCC}%
\end{align}
where $a$ and $b$ are constants. Also, $h_{mn}(x^m)$ is the two-dimensional metric
associated to the `angular section' $\Sigma_{k}$ , whose volume and curvature will be denoted by $V(\Sigma)$ and $2 k$, respectively.

By evaluating the general expressions (\ref{eq:Q_G}) and (\ref{eq:Q_phi}) for the 
above asymptotic boundary conditions, and
considering that the boundary is at $r=\infty$, we obtain the gravitational
contribution
\begin{equation}
\delta M_{G}=\frac{V(\Sigma)}{\kappa}[r\delta a+l\delta b+O(1/r)],
\label{eq:delta_mg}%
\end{equation}
and scalar contribution 
\begin{equation}
\delta M_{\phi}=\frac{V(\Sigma)}{l^{2}}[r\alpha\delta\alpha+\alpha\delta
\beta+2\beta\delta\alpha+O(1/r)]. \label{eq:delta_mphi}%
\end{equation}
Thus, we have the variation of the mass
\begin{equation}
\delta M=\frac{V(\Sigma)}{\kappa l^{2}}[r(l^{2}\delta a+\kappa\alpha
\delta\alpha)+l^{3}\delta b+\kappa(\alpha\delta\beta+2\beta\delta
\alpha)+O(1/r)]. \label{varmass}%
\end{equation}
The above expression for $\delta M$ is meaningful only in the case of
vanishing constraints. For the asymptotic conditions considered here,
$H_{\perp}=0$ implies
\begin{equation}
\frac{k+a}{\kappa}+\frac{\alpha^{2}}{2l^{2}}=0. \label{a}%
\end{equation}
In this way, the divergent piece in (\ref{varmass}) is removed and the
asymptotic variation of the mass takes a finite value
\begin{equation}
\delta M=\frac{V(\Sigma)}{\kappa l^{2}}[l^{3}\delta b+\kappa(\alpha\delta
\beta+2\beta\delta\alpha)]. \label{varmassfin}%
\end{equation}

To remove the variations from (\ref{varmassfin}) we need to impose boundary
conditions on the scalar field. In particular, the integration of  (\ref{varmassfin}) requires a
functional relation between $\alpha$ and $\beta$.  If we define 
$\beta=dW(\alpha)/d\alpha$, the mass of the spacetime is given by\footnote{The mass in (\ref{eq:mass1}) is defined up to a constant without variation. Since in four dimensions there is no Casimir energy, this 
constant is zero in order to fix a vanishing mass for the locally AdS spacetime.}
\begin{equation}
\label{eq:mass1}M=V(\Sigma)\left[  \frac{l b}{\kappa}+\frac{1}{l^{2}}\left(
\alpha\frac{dW(\alpha)}{d\alpha}+W(\alpha)\right)  \right].
\end{equation}
Indeed, we recover the result of \cite{Hertog:2004ns} (see, also,
\cite{Astefanesei:2008wz} for $5$-dimensional black holes). At this point, it
is important to emphasize that the coefficient of the first subleading term,
$\mu$, in the expansion (\ref{BCA}) of $g_{tt}$ does not appear explicitly in
the expression of the mass. In fact, in static spacetimes $g_{tt}$ is the lapse
function which is not a canonical variable and consequently, does not appear 
either in the constraints or in the surface terms. However, as we will see shortly, once we use the
equations of motion the situation will change.

Now, for a\textit{ given solution} with the required asymptotics, we have additional
information since not only the constraints are satisfied, but also the
equations of motion. The $E_{t}^{t}-E_{r}^{r}$ combination of the
Einstein-scalar field equations, which is not a constraint, is independent of
the scalar field potential and yields
\begin{equation}
E_{t}^{t}-E_{r}^{r}=\frac{2a+2k+\kappa\alpha^{2}l^{-2}}{r^{2}}+\frac
{-3\mu+3bl+4\kappa\alpha\beta l^{-2}}{r^{3}}+O(r^{-4})=0.
\end{equation}
The first term gives the same relation as the constraint $H_{\perp}=0$, but
the second one provides a relation containing $\mu$ and the parameters 
of the asymptotic expansions of $g_{rr}$ and the scalar field
\begin{equation}
bl=\mu-\frac{4}{3}\kappa\alpha\beta l^{-2}. \label{b}%
\end{equation}
Then, the mass can be written as 
\begin{equation}
M=V(\Sigma)\left[  \frac{\mu}{\kappa}+\frac{1}{l^{2}}\left(  W(\alpha
)-\frac{1}{3}\alpha\frac{dW(\alpha)}{d\alpha}\right)  \right].
\label{eq:mass11}%
\end{equation}
Therefore, there are only three situations when the mass reduces to
$M=\mu V(\Sigma)\kappa^{-1}$:

\begin{itemize}
\item $\alpha=0$: this is the usual Dirichlet boundary condition and ensures
asymptotic AdS invariance;

\item $\beta=0$: this is the Neumann boundary condition and also ensures
asymptotic AdS invariance;

\item $\beta=C\alpha^{2}$: this boundary condition\footnote{The fact that under this boundary condition the contribution of the scalar field vanishes was noticed in \cite{Amsel:2006uf} using a different approach.} corresponds to multi-trace deformations in the 
dual field theory \cite{Witten:2001ua} and is also compatible with the asymptotic AdS symmetry
 \cite{Henneaux:2006hk}.
\end{itemize}

It is important to emphasize that the relation between the Hamiltonian mass
and the parameter $\mu$ that appears in the expansion of $g_{tt}$ will allow
us to establish a clear relation with the AMD prescription.

\subsection{Logarithmic branch}

It is well known that a second order differential equation has two linearly
independent solutions. When the ratio of the roots of the indicial equation is
an integer, the solution may develop a logarithmic branch. This is exactly what 
happens when the scalar field saturates the BF bound, in which case the leading 
fall-off contains a logarithmic term  \cite{Henneaux:2004zi}.
However, we are interested in a scalar field with the conformal mass
$m^{2}=-2l^{-2}$. To obtain the logarithmic branch, a cubic term in the
asymptotic expansion of the scalar field potential is necessary \cite{
Henneaux:2006hk}
\begin{equation}
U(\phi)=-\frac{3}{l^{2}\kappa}-\frac{\phi^{2}}{l^{2}}+\lambda\phi^{3}%
+O(\phi^{4}),
\end{equation}
so that the fall-off of the scalar field to be considered is
\begin{equation}
\phi=\frac{\alpha}{r}+\frac{\beta}{r^{2}}+\gamma\frac{\ln(r)}{r^{2}}+O(r^{-3}),
\end{equation}
and the suitable asymptotic behaviour for the metric can be expressed as
\begin{align}
-g_{tt} &  =\frac{r^{2}}{l^{2}}+k-\frac{\mu}{r}+O(r^{-2}),\label{LB1}\\
g_{mn} &  =r^{2}h_{mn}+O(r^{-2}),\\
g_{rr} &  =\frac{l^{2}}{r^{2}}+\frac{al^{4}}{r^{4}}+\frac{l^{5}c\ln(r)}{r^{5}%
}+\frac{l^{5}b}{r^{5}}+O\left(  \frac{\ln(r)^{2}}{r^{6}}\right). \label{LB2}%
\end{align}
Using a similar procedure as in the previous section we get
\begin{equation}
M=\left[  \frac{lb}{\kappa}+\frac{1}{l^{2}}\left(  \alpha\frac{dW}{d\alpha
}+W(\alpha)+\alpha^{3}l^{2}\lambda\right)  \right]  V(\Sigma).
\end{equation}

To relate the mass with the first subleading term of $g_{tt}$ we use the
combination of the Einstein equations $E_{t}^{t}-E_{r}^{r}$ which yields
\begin{equation}
b=\frac{\mu}{l}-\frac{2\kappa\alpha\left(  \alpha^{2}l^{2}\lambda
+2\beta\right)  }{3l^{3}},%
\end{equation}
and then the mass becomes
\begin{equation}
M=\left[  \frac{\mu}{\kappa}+\frac{1}{l^{2}}\left(  W(\alpha)-\frac{1}%
{3}\alpha\frac{dW}{d\alpha}+\frac{1}{3}\alpha^{3}l^{2}\lambda\right)  \right]
V(\Sigma).
\end{equation}
Therefore, we obtain $M=\mu V(\Sigma)\kappa^{-1}$ only for $\alpha=0$ or
\begin{equation}
W(\alpha)=\alpha^{3}\left[  C+l^{2}\lambda\ln(\alpha)\right],
\end{equation}
which are the AdS invariant boundary conditions \cite{Henneaux:2006hk}.

\section{Ashtekar-Magnon-Das mass}

The AMD procedure \cite{Ashtekar:1984zz, Ashtekar:1999jx} is particularly
attractive because it can be straightforwardly applied to hairy black holes
(detailed applications related black hole physics can be found in
\cite{Acena:2013jya, Chen:2005zj}). The AMD conserved quantities are
constructed from the electric part of the Weyl tensor. First, consider a
conformally rescaled metric that is regular at the boundary
\begin{equation}
\tilde{g}_{\mu\nu}=\omega^{2}g_{\mu\nu},%
\end{equation}
where $g_{\mu\nu}$ is the asymptotically AdS metric of interest and $\omega$
has a zero of order one at infinity. $\tilde{g}_{\mu\nu}$ defines a conformal
structure at infinity since $\omega$ is defined up to a multiplication of a
regular function of the boundary coordinates. The central object of the AMD
prescription is the electric part of the Weyl tensor
\begin{equation}
\mathcal{E}_{\mu}^{\nu}=l^{2}\omega^{-1}n^{\alpha}n^{\beta}C_{\alpha\beta\mu
}^{\nu}, \label{AMD}%
\end{equation}
where $n_{\mu}=\partial_{\mu}\omega$ is the normal vector on the boundary and
$C_{\beta\alpha\mu}^{\nu}$ is the Weyl tensor of $\tilde{g}_{\mu\nu}$. Note
that all the objects in (\ref{AMD}) are intended to be calculated and
index-manipulated with the metric $\tilde{g}_{\mu\nu}$. The energy in both
cases, with or without logarithmic branches, is
\begin{equation}
M_{\textrm{AMD}}=\frac{l}{\kappa}\int_{\Sigma}\mathcal{E}_{tt}d\Sigma^{t}%
=\frac{\mu V(\Sigma)}{\kappa}. \label{AMD2}%
\end{equation}
It is now clear that AMD mass matches the actual mass of the spacetime,
defined by the Hamiltonian, only for AdS invariant boundary conditions.

\section{Discussion}

In this letter, we have computed the mass for asymptotically AdS 
configurations  endowed with a massive minimally coupled scalar 
field. It has been shown that the canonical generator associated to 
the time translation symmetry, i.e. the mass,  once evaluated using the
equations of motions, coincides with the coefficient of the first subleading term
of the lapse function only for boundary conditions that are compatible 
with the canonical realization of the local AdS symmetry at the boundary.

Additionally,  we have explicitly shown that the AMD mass provides
the right result, as defined by the Hamiltonian method, only for
boundary conditions that preserve the conformal invariance of the
boundary (and so of the dual theory)\footnote{We also have an
independent computation using the counterterms that supports this
claim \cite{noi}.}.

A conformal field theory (CFT) embedded in a curved spacetime background can
be characterized by the trace anomaly coefficients of the stress tensor. This
result applies to even dimensional CFTs because there is no trace anomaly in
odd dimensions. Since within the AdS/CFT duality a CFT has a gravity dual, one
expects that a dual gravitational computation will also account for the
Casimir energy. Indeed, this was explicitly proven in the so called
holographic renormalization \cite{Henningson:1998gx}. From a holographic point
of view, the quantum fluctuations contribute to the inertia in the boundary.
However, a flaw of the AMD method is that it has low accuracy in this regard,
namely it can not account for the Casimir energy. Since the dual field theory
is living on a $3$-dimensional sphere, there is no Casimir energy associated
and so, when the boundary conditions preserve the conformal symmetry, the AMD
prescription and the Hamiltonian method produce identical results. However,
when the boundary conditions do not preserve the conformal symmetry these
methods produce different results.

Let us comment now 
on the test particle motion in scalar hairy AdS spacetimes. This  could be related with
potentially observable effects. For the
clarity of the argument, let us make the discussion quantitative. Consider the
four-dimensional static asymptotically flat metric:
\begin{equation}
ds^{2}=-\left[  1-\frac{\mu}{r}+O(r^{-2})\right]  dt^{2}+\frac{dr^{2}}{\left[
1-\frac{m}{r}+O(r^{-2})\right]  }+r^{2}d\Omega^2 .\label{ST}%
\end{equation}
Note that we have parametrized differently the $O(r^{-1})$ term of $g_{tt}$
and $g_{rr}^{-1}$, respectively. When there is no contribution from the matter
fields, i. e. when the matter fields fall off fast enough at infinity, the Hamiltonian 
mass of the spacetime is
\begin{equation}
M=\frac{m}{2G}.\label{H}%
\end{equation}
Indeed, this is the case for massive scalar fields since they are exponentially suppressed in
asymptotically flat spacetimes, and consequently the field equations yield $\mu=m$. The motion of test particles on circular orbits is driven by the (mass) parameter $\mu$,  as is revealed  by the expression for the rate of revolution $\omega= d \phi / dt$ in a circular orbit at radius $R$ which, for a generic spherically symmetric spacetime,  is given by
\begin{equation} \label{w}
\omega^2= -\frac{1}{2 R} \frac{d g_{tt}}{d r} |_{r=R} \,\, .
\end{equation}
Then, for the asymptoticaly flat metric (\ref{ST}), we have $ \omega^2= \mu /(2 R^3)$ when $R$ is large. Therefore, this parameter can be interpreted as the gravitational mass that
generates the gravitational field responsible for the test
particles' motion. If one interprets the Hamiltonian mass as the
inertial mass of the system, then, in agreement with the equivalence
principle, it is not surprising that $m=\mu$.

It is instructive to contrast the previous result with the circular geodesic
motion on an asymptotically AdS spacetime in the presence of a massive 
scalar field, described by the metric
\begin{equation}
ds^{2}=-\left[  \frac{r^{2}}{l^{2}}+1-\frac{\mu}{r}+O(r^{-2})\right]
dt^{2}+\frac{dr^{2}}{\left[  \frac{r^{2}}{l^{2}}+1-\frac{m(r)}{r}%
+O(r^{-2})\right]  }+r^{2}d\Omega^2,\label{AdS}%
\end{equation}
where $m(r)$ grows slower than $r^3$ \cite{Henneaux:2006hk}. In this case, 
the rate of revolution reduces to $l^{-2}+\mu /(2 R^3)+O(R^{-4})$. 
As is expected, apart from the parameter $\mu$, the motion is driven also 
by the cosmological constant. Considering the latter constant as a fundamental one, the
motion is addressed only by $\mu$.  However, when the backreaction of scalar
fields is taken into account, the mass matches $\mu /(2G)$ only for the AdS invariant boundary conditions discussed in the previous section. This suggests that an issue arises from the interpretation of $\mu$ as the gravitational mass when the boundary condition on the scalar field breaks the conformal symmetry. 

One obvious extension of this work is an application to higher dimensional
scalar hairy black holes \cite{Acena:2013jya, Feng:2013tza}. Also, one can
study the charged hairy black holes and their extremal limits. In the extremal
limit, the attractor mechanism plays the role of a no-hair theorem
\cite{Astefanesei:2007vh} in the sense that the moduli are fixed at the
horizon and the near horizon geometry is universal. The moduli flow is
interpreted as an RG flow and it will be interesting to compare the charges
computed at the horizon \cite{Astefanesei:2011pz, Astefanesei:2010dk} with the charges computed in
the boundary. In this way, one can understand better the role of the hair
(scalar degrees of freedom living outside the horizon) for black hole physics.

\vskip 1cm

\section*{Acknowledgments}

We would like to thank Robert Brandenberger and Kurt Hinterbichler for
interesting discussions. This work has been partially funded by the  Fondecyt 
grants 11121187, 1120446, 1121031,  and 1130658.
The Centro de Estudios Cient\'{\i}ficos (CECs) is funded by the Chilean Government
through the Centers of Excellence Base Financing Program of Conicyt.


\end{document}